\documentclass{jetpl}
\twocolumn
\usepackage{graphicx}
\usepackage{bm}

\lat

\title{On superconducting and magnetic properties of iron-oxypnictides}

\rtitle{On superconducting and magnetic properties of iron-oxypnictides}

\sodtitle{On superconducting and magnetic properties of iron-oxypnictides}

\author{V.\,Barzykin$^{\dagger}$,
L.\,P.\,Gorkov$^{*,\ddagger}$\/\thanks{e-mail: gorkov@magnet.fsu.edu}}

\rauthor{V.\,Barzykin, L.\,P.\,Gorkov}

\sodauthor{Barzykin, Gorkov}

\address{
$^{\dagger}$ Department of Physics and Astronomy, University of Tennessee,
Knoxville, TN  37996 USA \\~\\
$^*$National High Magnetic Field Laboratory,
Florida State University,
1800 E. Paul Dirac Dr., Tallahassee, Florida 32310 USA\\~\\
$^{\ddagger}$ L.D. Landau Institute for Theoretical Physics RAS,
142432 Chernogolovka, Russia}

\dates{11 June 2008}{*}

\abstract{Pairing symmetry in oxypnictides, a new
family of multiband high-$T_c$ superconductors, is partially imposed by the 
positions of multiple Fermi pockets, which itself can give rise to new 
order parameters, such as $s_{+,-}$-states or the state of $d_{x^2-y^2}$ symmetry. 
Other pairing states may appear on small pockets for long range interactions, but 
they are expected to be sensitive to defects. We identify the competing antiferromagnetic order with the 
triplet exciton transition in the semi-metallic background and discuss whether its
coexistence with superconductivity explains the doping dependence of $T_c$.}


\begin{document}

\maketitle

Recently discovered superconductors (SC) among the series of iron-oxypnictides\cite{Kamihara} with 
unexpectedly high SC transition temperatures (T$_c$) reveal a tantalizing resemblance to 
the cuprates: the layered tetragonal structure, an antiferromagnetic (AFM) order in the 
normal state and an insulating or semi-metallic behavior for the parent stoichiometric 
materials with sharp increase in metallicity and the value of T$_c$ for both  electron- and hole-
doped materials\cite{HHWen}.

	While the experimental progress is still hindered by the shortage of single crystals, 
impatience to sort out the physics that governs the puzzling magnetic and electronic 
behavior in these new superconductors generated an avalanche of theoretical attempts to 
analyze the new materials from different perspectives, although, most often, using 
numerical methods. Currently the discussion of physical mechanisms is 
difficult, in part, because quantitative results often contradict each other. 
Below we address the phenomena in oxypnictides in a more phenomenological 
manner, using, at the same time, a number of qualitative features that emerged from previous 
analyses and of which a consensus was achieved. 

We consider possible symmetries of SC order parameters 
in terms of the SC symmetry classes\cite{VG}. 
The realization of symmetries in a multiband model for oxypnictides leads to 
interesting implications that depend on the interactions' details. We argue 
that the spin density wave (SDW) in these materials is the triplet excitonic state known for low carrier systems 
since the 60-s\cite{mott,knox} (see Ref.\cite{HR} for review). We discuss the coexistence
of SDW and SC and the dependence of $T_c$ on doping. 

      The reasons why T$_c$ strongly varies among REOFeAs (RE stands for rare earth)\cite{Kamihara,XHChen,GFChen,ren1}, 
from a few K in the stoichiometric LaOFeAs up to 55K for RE=Sm\cite{ren2}, remain poorly understood. For instance, T$_c$
is not sensitive to the choice of RE in the cuprates. 
It was suggested empirically\cite{WLu} that T$_c$ increases with decreased ionic radii of the  
rare earth due to the inner chemical pressure. Band structure calculations\cite{nekrasov}, however, found 
no dependence on the RE-element. Therefore, we mainly discuss below the doping dependence of T$_c$.  

	 There is also no consensus about the applicability of the Fermi liquid concepts to 
oxypnictides. The importance of strong correlations was emphasized in Ref.\cite{haule}.
To account for the experimental magnetic phase diagram of the parent compound, 
LaOFeAs\cite{mook,McGuire}, localized moments on the d-orbital Fe-ions were postulated 
in Refs. \cite{abrahams,yildirim,kivelson}.    
Meanwhile,
band structure calculations\cite{nekrasov,DJSingh,kuroki} predict up to five small 2D Fermi surface (FS) sheets in 
LaOFeAs; the material would then constitute a low carrier 2D ionic metal that, within the 
limits of accuracy of the method, shows a proximity to either ferro- or AFM- 
instabilities. According to Ref.\cite{Mazin}, LaOFeAs must be itinerant. The tendency to 
AFM state ascribed to ``nesting'' between different FS sheets.
 
	The metallic behavior\cite{Kamihara,Dong} above T$_c$ and the very onset of SC present 
a direct proof of an itinerant character for the carriers in the oxypnictides. 
Therefore, FS sheets \textit{must} 
exist, at least at finite doping (see, e.g.,Ref.\cite{kuroki}). In Ref.\cite{DJSingh} these FS, as it was mentioned 
above, were obtained from the Density Functional calculations. Note 
that all numerical methods become especially vulnerable near the chemical potential, $E_F$, 
and often lead to conflicting results for the density of states (DOS), such as asymmetries 
in DOS at $E_F$ that would contradict observations of SC for both electron\cite{Kamihara} and hole\cite{HHWen} 
doping\cite{ZPYin}.

\begin{figure}
\includegraphics[width=3.375in]{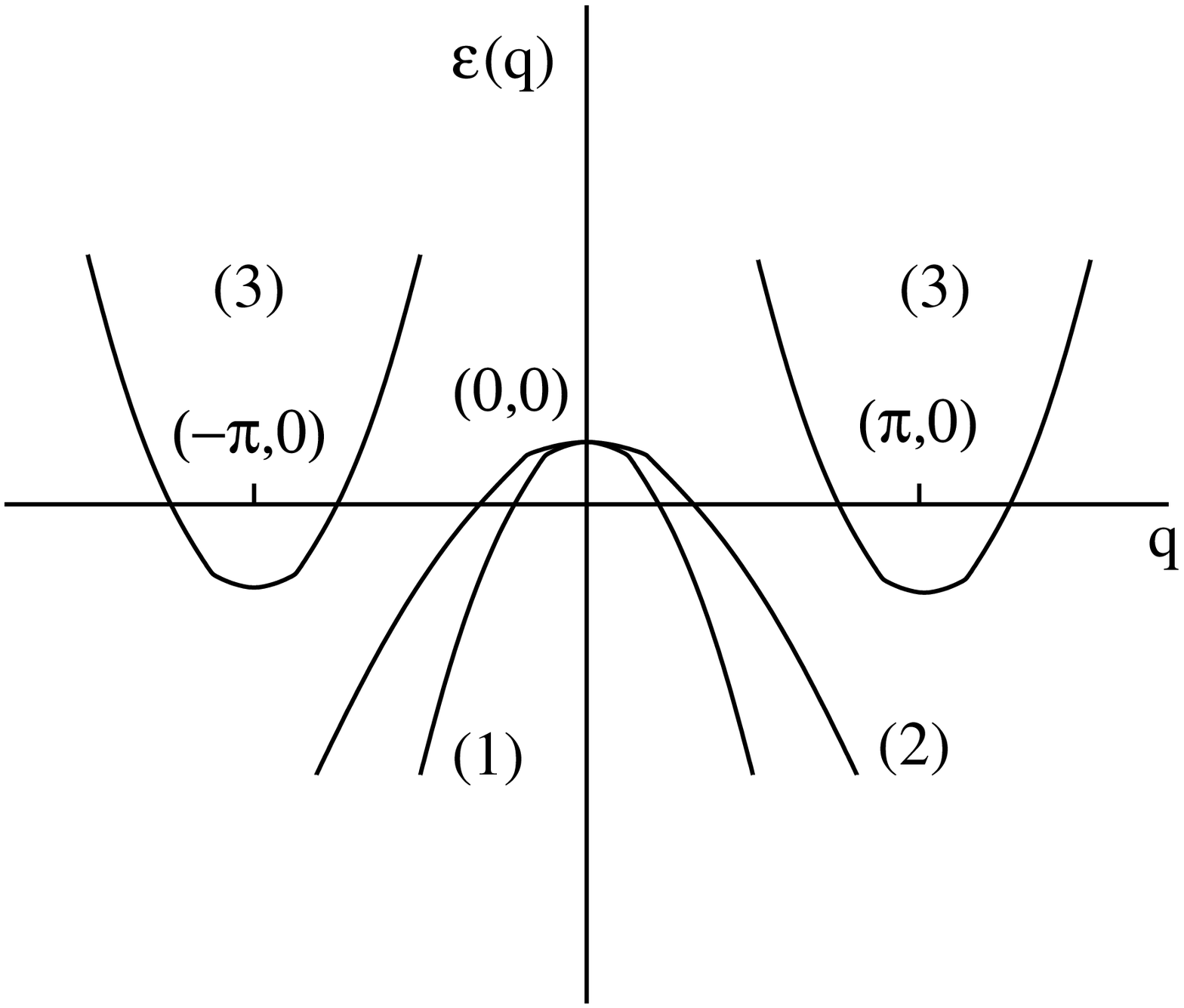}
\includegraphics[width=3.375in]{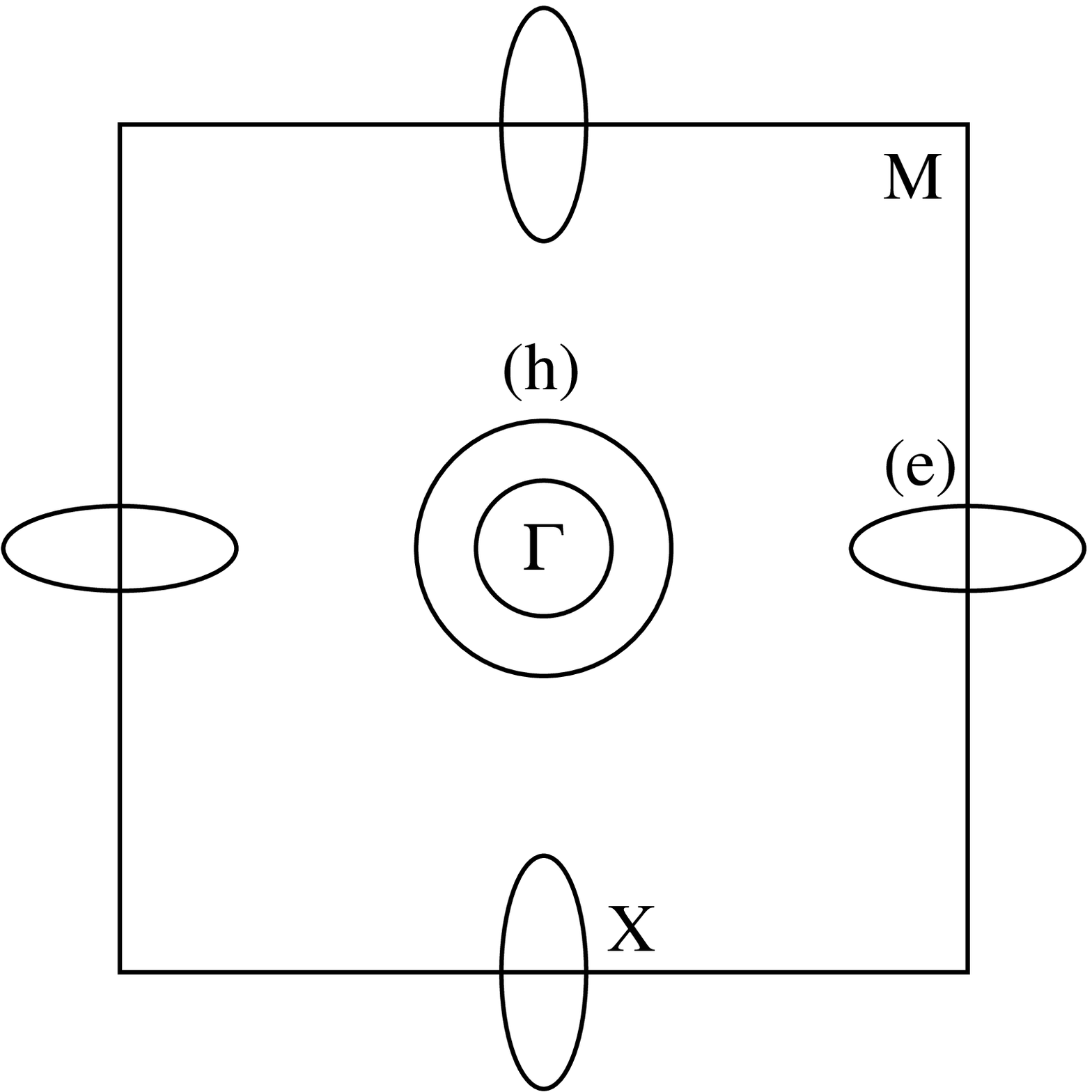}
\caption{Schematic electronic spectrum (a) and Fermi surfaces (b) of LaOFeAs in the unfolded 
Brillouin Zone. Two different hole Fermi surfaces form around the $\Gamma$ point, while the 
electron pockets appear at the $X$-points.}
\label{fig1}
\end{figure}

We accept below the model for the electronic structure of the oxypnictides with the FS topology 
that practically coincides with the one in Ref.\cite{DJSingh,Mazin}, except that we omit the small 3D hole 
pocket at the Z-point in the large Brillouin Zone (BZ) (1Fe/cell). In Fig. \ref{fig1} there are two hole 
pockets in the center of the BZ that appear as a result of the degeneracy of the spectrum at the 
$\Gamma$- point\cite{nekrasov,kuroki}. Two electron pockets on the sides of the square, 
$(0,\pi)$ and $(\pi, 0)$, have a two-fold anisotropy.

	Such FS structure corresponds, at the stoichiometric composition, to the 
compensated semi-metal, with equal volumes occupied by electrons and holes. Electron doping 
would diminish sizes of the hole-pockets, and vice versa. Judging from 
experiments\cite{Kamihara,HHWen}, DOS on both sides of the Fermi level, $\nu(E_F)$, 
is approximately the same and \textit{stays  constant} in its vicinity, reflecting
the two-dimensionality of FS. This is not the case in band structure 
calculations\cite{nekrasov,haule,DJSingh}. On the other hand, if DOS \textit{were 
a constant, the transition temperature, T$_c$, could not vary with concentration} 
in the BCS-like fashion: 
\begin{equation}
T_c(x) \propto exp(-1/V \nu(x,E_F)).
\label{TcBCS}
\end{equation}
It is known experimentally that SC competes with AFM: no magnetic order has 
been observed either in the normal or SC state for La(O$_{1-x}$F$_x$)FeAs ($T_c = 26 K$)\cite{YQui}. 
However, if the SDW and SC orders did coexist, this could provide a natural mechanism for a variable DOS near 
$E_F$\cite{Dong}.  

For what follows it is helpful to consider
the characteristic energy scales of oxypnictides. These are as follows: 
the Debye temperature is $\omega_D = 282 K$\cite{Dong}; the energies above and below $E_F$ spanned by the 
iron d-orbitals are in the range $\Delta E \sim 2 eV$, while the electron- and hole- pockets' depths are 
of the order of $0.1-0.2eV \sim 2000K \ll \Delta E$\cite{nekrasov,DJSingh,kuroki}; the 
temperature of the SDW singularity in the parent compound LaOFeAs is $T_{SDW} \sim 150 K$\cite{mook}, and 
the corresponding DOS energy gap, as seen in the infrared reflectance\cite{Dong} is small: $\sim 150-350 cm^{-1}$.
With T$_c$ ranging from $\sim 4 K$ to $55 K$, it seems possible to analyze the properties of oxypnictides 
at the mean field level. 

	We start with the symmetry analysis for the SC order parameters on 
multiple FS's in Fig. \ref{fig1} in the absence of SDW. Some results are also rigorously applicable 
when the SDW and either the hole- or electron- pockets are fully eliminated by the proper doping.
Note that the pockets' areas in Fig.\ref{fig1} are rather small 
(of order 10 \% or less of BZ, e.g.,see Ref.\cite{DJSingh}).
Let $\Delta_i (\bm{p})$ be the SC parameter on the $i^{th}$ FS. The self-consistency relation that defines 
$\Delta_i (\bm{p})$ through the anomalous Gor'kov functions, $F_i(\omega_n, \bm{p})$, is of the form:
\begin{equation}
\Delta_i (\bm{p})= T \sum_{j; \omega_n} \int V^{i,j}(\bm{p}-\bm{p'}) d \bm{p'} F_j(\omega_n, \bm{p'})
\label{gapeq}
\end{equation}
where $V^{i,j}(\bm{p}-\bm{p'})$ is the Fourier component of the interaction. If the pockets are small, 
the momentum transfer, $\bm{p}-\bm{p'}$, inside each pocket is small as well, 
so that one can replace 
$V^{i,j}(\bm{p}-\bm{p'})$  with $V^{i,j}(0)$, if, as we discuss below, the 
interaction is short range. This, of course, favors momentum independent gaps and, hence, a singlet pairing.

The interaction Hamiltonian for four separate Fermi surfaces can be written in the BCS-like form 
(compare, for example, with  Ref. \cite{ABG}):

\begin{equation}
H_{int} =  \frac{1}{2}\, \sum_{\bm{p},\bm{p'}} 
\sum_{i j \sigma \sigma'} \tilde{V}^{i,j}_{\bm{p},\bm{p'}} 
a^{\dagger}_{i \sigma \bm{p}} a^{\dagger}_{i \sigma' - \bm{p}}
a_{j \sigma' \bm{p'}} a_{j \sigma \bm{-p'}},
\label{elec}
\end{equation}
and
\begin{equation}
\tilde{V}^{i,j}_{\bm{p}, \bm{p'}} = V^{i,j}(\bm{p}-\bm{p'})w^*_i (\bm{p})w^*_i (-\bm{p}) w_j(\bm{p'}) w_j(-\bm{p'}),
\end{equation}
where $w_i(\bm{p})$ are the periodic Bloch functions. The anomalous Gor'kov functions 
$F$,$F^{\dagger}$ are defined in the real space. Correspondingly, the wave functions of the Cooper pair are 
proportional to $\langle \Psi_i(\bm{r}) \Psi_j (\bm{r'}) \rangle$, where $\Psi_i(\bm{r})$ is the full
field operator. Therefore, $\Delta_i (\bm{p})$
differ from just the averages $\langle a_{i, \bm{p}} a_{i,-\bm{p}} \rangle$, where $\bm{p}$ is 
now the quasi-momentum, by the factors $w_i(\bm{p})$ and $w_i(-\bm{p})$. The derivation of the Bloch 
functions  and of the energy spectrum near the double-degenerate $\Gamma$-point will be given 
elsewhere.

	To apply the approach of Ref.\cite{VG} to the multiband model in Fig.\ref{fig1}, it is necessary to account for 
the fact that the transformations of the point group $D_4$ may interchange the electronic FSes. 
After linearization in $\Delta_i$, Eq.(\ref{gapeq})  
determines the transition temperature for each irreducible representation. 
To enumerate all possible symmetries, we apply the transformations of the 
$D_4$  group to the column of $\{\Delta_{h1}, \Delta_{h2}, \Delta_{eX}, \Delta_{eY}\}$.
After calculating characters we find three identical  representations 
and the one belonging to the non-trivial class, $B_2$. 
The latter constitutes the $d_{x^2-y^2}$ state which, if one neglects the momentum dependence of $V^{i,j}(\bm{p}-\bm{p'})$, 
 signifies a gapless state on the hole pockets in Fig. \ref{fig1}. (For momentum dependent interactions, however, the hole 
pockets become weakly gapped with zeroes along the diagonals in Fig.\ref{fig1}.)
Recent high resolution PES experiments\cite{XJia} have indeed detected gapless hole FSes at the $\Gamma$-point. 

To account for all the features of the band structure in Fig. \ref{fig1}, we perform this analysis in a 
more explicit manner. The interaction between the electrons forming a Cooper pair 
on Fermi pockets shown in Fig. \ref{fig1} takes the following form:
\begin{equation}
V=\begin{pmatrix} 
u & u & t & t \cr
u & u & t & t \cr 
t & t & \lambda & \mu \cr
t & t & \mu & \lambda 
\end{pmatrix}.
\label{mmatr}
\end{equation}
Here $\lambda = V^{eX,eX} = V^{eY,eY}$ is the interaction on the same electron pocket at the $X$-point, 
$\mu = V^{eX,eY}$
couples electron pockets at $(\pi, 0)$ and $(0, \pi)$,  
$u = V^{h1,h1}=V^{h2,h2}=V^{h1,h2}$ characterizes the BCS interactions for electrons on two different hole Fermi surfaces
at the  $\Gamma$-point, while $t = V^{h,eX}=V^{h,eY}$ is the coupling that connects the Cooper pairs at the $X$-points
and the $\Gamma$-point. The form of the interaction matrix $V$ for the two hole pockets at 
the $\Gamma$-point follows from the degeneracy of the bands at the $\Gamma$-point.  

Solution of the linearized gap equation,
\begin{equation}
\Delta_{i}  =   \sum_{j} \bar{V}^{i,j} \Delta_{j} \ln{\frac{2 \gamma \bar{\omega}}{\pi T_c}\,}\,,
\label{gapeqlin}
\end{equation}
where $\bar{\omega}$ is the cut-off in the Cooper logarithm and
\begin{equation}
\bar{V}^{i,j} \equiv  - \frac{1}{2}\, V^{i,j} \nu_{j},
\label{intmatr}
\end{equation}
determine the $T_c$ values for the order parameters of different symmetries. 
Introducing effective coupling $g$,
\begin{equation}
T_c = \frac{2 \gamma \bar{\omega}}{\pi}\,e^{- 2/g}, 
\label{TC1} 
\end{equation}
we find three different solutions:

1) For the non-trivial, $d_{x^2 - y^2}$ symmetry - the energy gap on different $X$-pockets changes sign:

\begin{equation}
\Delta_1 = \Delta_2 = 0, \ \ \Delta_3 = - \Delta_4 = \Delta,
\end{equation}

\begin{equation}
g = (\mu - \lambda) \nu_3. 
\label{dwve}
\end{equation}

2) The two solutions that can belong to the so-called $s_{+}$, $s_{-}$ states, where the energy gap at 
the $X$-points have the same sign, while at the $\Gamma$-point 
it may have a different sign, with

\begin{eqnarray}
\label{swaveg}
2 g_{+,-} &=& - u (\nu_1 + \nu_2) - (\lambda + \mu) \nu_3 \pm  \\
& & \sqrt{(u (\nu_1 + \nu_2) - (\lambda + \mu) \nu_3)^2 + 8 t^2 \nu_3 (\nu_1 + \nu_2)} \nonumber  
\end{eqnarray}
and
\begin{equation}
\Delta_1 = \frac{\nu_1 \Delta_2}{\nu_2} = \kappa \Delta , \ \ \Delta_3 = \Delta_4 = \Delta,
\label{swaveord}
\end{equation}
where $\kappa^{-1} = - (g_{+,-} + u (\nu_1 + \nu_2))/(t \nu_1)$.

	In Ref. \cite{Mazin} it was assumed that the strongest interactions may be due to the SDW channel.
In our notations it would mean large $V^{e-h}$ matrix elements in Eqs. (\ref{elec}),(\ref{mmatr}),(\ref{gapeqlin}).
The "$s^{+(-)}$" superconductivity\cite{Mazin} for gaps on the electron and hole FS 
indeed follows from Eqs(\ref{swaveg}),(\ref{swaveord}) in this limit.
(Such a state does not follow from the symmetry arguments).

   We turn now to the changes in Fig.\ref{fig1} introduced by the presence of  
spin density wave (SDW). The AFM phase itself can be described in terms 
of the local iron moments provided the exchange integrals are such that the (nn) exchange parameter, 
$J_1$, is smaller than the (nnn) integral, $J_2$. This relation leads to the AFM state with the correct structure 
vector $\bm{Q}= (0, \pi)$ or $(\pi,0)$\cite{abrahams,yildirim,kivelson}.    
For $J_1$, $J_2$ experiments
would then give an estimate $J \sim 150-250K$\cite{kivelson,mook}. (Theoretical estimates vary 
strongly from $\sim 0.5 eV$\cite{abrahams} to $\sim 50-100K$\cite{haule}). 

	Exchange integrals in the Hamiltonian for local spins determine the Ne\`{e}l 
temperature, $T_N$. However, in this case AFM introduces no new scale in the momentum space:
the magnon spectrum extends over the whole BZ. Therefore, 
the same arguments as above apply to contributions into the $e-e$ interactions 
that originate from the degrees of freedom of local spins : 
if FS pockets are of small size, the dependence on momentum transfer in the 
scattering matrix element of two electrons inside the same pocket
may be omitted.

	SDW in oxypnictides is commonly attributed to the strong nesting in the literature,
 i.e., the congruency in the FS shapes between electron and hole pockets in Fig.\ref{fig1} 
(see e.g. in Refs. \cite{McGuire,DJSingh,Mazin,raghu}). 
The generalized magnetic susceptibility, $\chi( \bm{q})$, numerically calculated in 
Refs.\cite{DJSingh,Mazin},  demonstrates indeed the strongly peaked character at $\bm{Q} = (0, \pi)$. 
Among advantages of this scenario,  in the first place, is that it leads naturally to the 
observed AFM structure vector\cite{mook}. 
Note, however, that the vector $\bm{q}=\bm{Q}$, in our opinion, does not justify the nesting scenario.
Indeed, the 
electron pocket contains only half of the carriers' number at the $\Gamma$ - point (for the parent compound), 
and, in addition, is anisotropic. Therefore, we suggest that the numerical 
results\cite{DJSingh,Mazin} do not distinguish between the nesting scenario 
and the formation of the exciton phase\cite{mott,knox,HR}, which comes about, 
first of all, due to the Coulomb interactions between the electron and the hole bands. 
(In the main approximation the instability is degenerate with respect to
formation of singlet or triplet excitons\cite{HR}). 

	For the AFM superstructure with vector $\bm{Q} = (0,\pi)$, 
the two hole-like bands and one electron band in Fig.\ref{fig1} overlap, leading to a partially gapped
spectrum because of non-congruency of pockets. The second electron pocket 
remains untouched.  
Assuming the interactions discussed above, it is natural to expect smaller T$_c$ due to
partially gapped spectra. As SDW is gradually lifted with doping, the T$_c$-value is expected to increase.
This process, at first glance, could provide a mechanism for the variation of $T_c$ at low doping.	

No rigorous theoretical results are available for the excitonic transition beyond the Hartree-Fock 
approximation (see, e.g., in Ref.\cite{HR}). Qualitatively, dependence
of such a phase on doping is as follows. 
The value of $T_N$ or, better, the gap\cite{Dong} provides an estimate of the energy gain at the phase formation. 
SDW would be 
destroyed when the change in the energy due to carriers poured 
into the electron or hole pockets, becomes of the same order.
 Rough estimates for the carriers' concentration 
 give realistic values for $x$ a few percent (the estimate
is sensitive to the numerical results for the effective mass
that vary considerably\cite{nekrasov,DJSingh,ZPYin}).

	In LaOFeAs the SDW -anomaly disappears already at 3 \% F-doping, while T$_c$ 
continues to increase with doping up to 28K at $6 \sim 8$ \% F-content\cite{Dong}.
The AFM feature seen at 150K in pure NdOFeAs\cite{GFChen1} rapidly 
disappears at the oxygen doping\cite{ren3} below the $T_c$-maximum (a sudden
increase in $T_c$\cite{ren3} can be a hallmark of a change in the ground state).  
One, hence, concludes that SDW and SC change 
differently and independently with doping and the SDW is destroyed faster. 

	The above brings us back to the problem that in the BCS-like Eq.(\ref{TcBCS}) for T$_c$ the two-dimensional DOS 
does not vary with the carriers' concentration and the concentration dependence of T$_c$, hence, 
is to reflect new mechanisms. It is possible that doping concentrations up to 15 \% are not so small
 and, in view of the ionic character of the coupling between oppositely charged layers, 
produces strong effects on the lattice\cite{WLu,ren3}. Another, although a more exotic, 
interpretation in frameworks of rigid bands, could signify 
an enhancement  in the interactions strength. 
No changes in the phonon spectrum have been reported\cite{Dong}. The Coulomb screening in 2D in the Thomas-Fermi 
limit does not depend on carriers concentrations  either. (Actual Coulomb screening acquires three-dimensional
character through charges in the adjacent planes).  As to AFM fluctuations (e.g., in \cite{Mazin}), 
in view of the above, one may expect that AFM fluctuations are already significantly suppressed 
in this doping range.  

	 The low sensitivity of T$_c$ to defects favors the gapped SC order.  
(The classes that allow nodes
along a FS, of course, should be same as in Ref.\cite{VG} and have been recently discussed anew, 
e.g., in Refs.\cite{ZHWong,YWan}). The observation of gapless FS at the $\Gamma$-point\cite{XJia}
points unambiguously in favor of the $d_{x^2-y^2}$ SC order for the electronically doped materials. 
In other words, at high enough doping level SC pairing is not electron-hole symmetric. 
For the "d-wave" pairing in Eq (\ref{dwve}) we can assume negative $\lambda$( phonons) and positive $\mu$ 
(Coulomb repulsion).

	In summary, we have enumerated possible symmetries of SC order  for the FS topology now
broadly accepted for the oxypnictides in the literature. Small sizes of FS in the frameworks of a semi-metallic 
picture allowed to narrow  significantly the number of SC states.  Among the three SC states there exists the symmetry
class analogous to the popular $d_{x^2-y^2}$ pairing in the cuprates. Its realization in oxypnictides predicts 
the whole ungapped internal (hole) pockets, in excellent agreement with the recent high resolution PES 
experiments\cite{XJia}. We identified onset of the AFM order in the parent materials with transition into the 
triplet exciton state caused by Coulomb interaction in this family of semi-metallic compounds.
While at low doping AFM affects the band energy spectrum and the resulting variation in DOS can explain
the dependence of T$_c$ on concentration, experimentally T$_c$ continues
to increase even after AFM is destroyed. 
Our results show that SC for electron and hole doping have a different symmetry.  
We have also discussed possible mechanisms for the increase of $T_c$  with doping. 

The authors are thankful to Z. Fisk, M. Sadovskii, and G.Teite'lbaum   
for helpful discussions. This work was supported by NHFML through the NSF
Cooperative agreement No. DMR-008473 and the State
of Florida.

\end{document}